%% file: main.tex
\documentclass[sigconf]{acmart}
\usepackage[mathscr]{eucal}
\usepackage{amsmath,amsthm,bm}
\usepackage{enumitem}
\usepackage{multirow}
\usepackage{booktabs}
\usepackage{scalerel}
\usepackage[multiple]{footmisc}
\usepackage{pgfplots}
\usepackage{pifont}
\usepackage{algorithm} %,algorithmic
\usepackage{algpseudocode}
\usepackage{tabularx}
\usepackage{footnote}
\usepackage{balance}
\usepackage{graphicx}
\usepackage{nicefrac}
\usepackage{subcaption}
\usepackage{textcomp}

\makeatletter
\def\BibTeX{{\rm B\kern-.05em{\sc i\kern-.025em b}\kern-.08em
    T\kern-.1667em\lower.7ex\hbox{E}\kern-.125emX}}

\newcommand{\multiline}[1]{%
  \begin{tabularx}{\dimexpr\linewidth-\ALG@thistlm}[t]{@{}X@{}}
    #1
  \end{tabularx}
}

\pgfplotsset{compat=1.5.1}
\def\addlegendimage{\csname pgfplots@addlegendimage\endcsname}
\usetikzlibrary{patterns}
\usetikzlibrary{pgfplots.groupplots}
\usetikzlibrary{
  pgfplots.colorbrewer,
}
\pgfplotsset{
  cycle list/.define={my marks}{
    every mark/.append style={solid,fill=\pgfkeysvalueof{/pgfplots/mark list fill}},mark=*\\
    every mark/.append style={solid,fill=\pgfkeysvalueof{/pgfplots/mark list fill}},mark=square*\\
    every mark/.append style={solid,fill=\pgfkeysvalueof{/pgfplots/mark list fill}},mark=triangle*\\
    every mark/.append style={solid,fill=\pgfkeysvalueof{/pgfplots/mark list fill}},mark=diamond*\\
  },
}
\newcommand{\xmark}{\ding{55}}
\DeclareMathSymbol{\mh}{\mathord}{operators}{`\-}
\definecolor{aa}{rgb}{0.19,0.55,0.91}
\definecolor{bb}{rgb}{0.9,0.17,0.31}
\definecolor{cc}{rgb}{0.514,0.325,0.831}
\definecolor{dd}{rgb}{0.12, 0.3, 0.17}
\definecolor{ee}{rgb}{1.0, 0.75, 0.0}
\definecolor{ff}{rgb}{0.01, 0.28, 1.0}
\definecolor{gg}{rgb}{0.65, 0.16, 0.16}

\AtBeginDocument{%
  \providecommand\BibTeX{{%
    Bib\TeX}}}

\copyrightyear{2026}
\acmYear{2026}
\setcopyright{cc}
\setcctype{by}
\acmConference[WWW '26]{Proceedings of the ACM Web Conference 2026}{April 13--17, 2026}{Dubai, United Arab Emirates}
\acmBooktitle{Proceedings of the ACM Web Conference 2026 (WWW '26), April 13--17, 2026, Dubai, United Arab Emirates}
\acmPrice{}
\acmDOI{10.1145/3774904.3792859}
\acmISBN{979-8-4007-2307-0/2026/04}

\newcommand{\ie}{{\it i.e.}}
\newcommand{\eg}{{\it e.g.}}
\newcommand{\rec}{{\textsc {\textsf ReCon}}}

\begin{document}

\title[Improving the Accuracy of Community Detection on Signed Networks\\via Community Refinement and Contrastive Learning]{Improving the Accuracy of Community Detection on Signed Networks via Community Refinement and Contrastive Learning} 

\author{Hyunuk Shin}
\authornote{These authors contributed equally as co-first authors.}
\affiliation{%
  \institution{Chungbuk National University}
  \city{Cheongju}
  \country{South Korea}
}
\email{shu914@cbnu.ac.kr}

\author{Hojin Kim}
\authornotemark[1]
\affiliation{%
  \institution{Chungbuk National University}
  \city{Cheongju}
  \country{South Korea}
}
\email{khojin.01@cbnu.ac.kr}

\author{Chanyoung Lee}
\authornotemark[1]
\affiliation{%
  \institution{Chungbuk National University}
  \city{Cheongju}
  \country{South Korea}
}
\email{csdud77@cbnu.ac.kr}

\author{Yeon-Chang Lee}
\affiliation{%
 \institution{Ulsan National Institute of Science and Technology (UNIST)}
 \city{Ulsan}
  \country{South Korea}
}
\email{yeonchang@unist.ac.kr}

\author{David Yoon Suk Kang}
\authornote{Corresponding author.}
\affiliation{%
  \institution{Chungbuk National University}
  \city{Cheongju}
  \country{South Korea}
}
\email{dyskang@cbnu.ac.kr}

\renewcommand{\shortauthors}{Hyunuk Shin, Hojin Kim, Chanyong Lee, Yeon-Chang Lee, \& David Yoon Suk Kang}

\begin{abstract}
Community detection (CD) on signed networks is crucial for understanding how positive and negative relations jointly shape network structure.
However, existing CD methods often yield inconsistent communities due to noisy or conflicting edge signs.
In this paper, we propose \textsc{\textsf{ReCon}}, a model-agnostic post-processing framework that progressively refines community structures through four iterative steps: (1) \textit{structural refinement}, (2) \textit{boundary refinement}, (3) \textit{contrastive learning}, and (4) \textit{clustering}.
Extensive experiments on eighteen synthetic and four real-world networks using four CD methods demonstrate that \textsf{ReCon} consistently enhances community detection accuracy, serving as an effective and easily integrable solution for reliable CD across diverse network properties.
\end{abstract}





\maketitle

\input{s1}

\input{s2}
\input{s4}

\input{s5}
\input{s6}

\begin{acks}
The work of David Y. Kang was supported by the National Research Foundation of Korea (NRF) grant funded by the Korea government (MSIT) (No. RS-2025-25435830).
The work of Yeon-Chang Lee was supported by Institute of Information \& communications Technology Planning \& Evaluation (IITP) grant funded by the Korea government (MSIT) (No.RS-2020-II201336, Artificial Intelligence Graduate School Program (UNIST))
\end{acks}

\bibliographystyle{ACM-Reference-Format}
\bibliography{sample-base2}










\end{document}

%% file: s1.tex
\section{Introduction}~\label{s1}
\noindent \textbf{Background.} 
The emergence of signed networks—which include both positive and negative edges—has enabled a deeper understanding of complex relational patterns among nodes~\cite{kim24:cikm, kan23:tkde}. 
By explicitly modeling both cooperative and antagonistic relations, these networks provide a more comprehensive view of social and relational structures. 
They are widely observed in web-based platforms such as social media, online forums, and review systems, where users express agreement (\ie, positive edges) or disagreement (\ie, negative edges) toward one another or shared content~\cite{les10:chi}. 
In such networks, identifying groups of similar nodes (\ie, communities) is crucial for understanding how positive and negative relations jointly shape the global organization and dynamics. 
Community detection (CD) on signed networks seeks to uncover these communities by analyzing two key connectivity patterns~\cite{yan17:tkde, kan23:tkde}: 
(1) most intra-community edges are positive, and (2) most inter-community edges are negative.

\vspace{1mm}
\noindent \textbf{Motivation.}
Despite notable advances, existing CD methods remain highly vulnerable to performance degradation in the presence of noisy edges, \ie, edge signs that conflict with underlying community structures.~\cite{kan23:tkde}.
In real-world datasets, user interactions and sentiment annotations are often ambiguous, leading to incorrectly assigned edge signs that distort the underlying relational structure. 
Since most CD methods assume clean and reliable signs, they are inherently susceptible to such noise. 
Although this issue has been conceptually recognized in prior study~\cite{kan23:tkde}, we empirically validate it through controlled experiments, showing that as the proportion of noisy edges increases, the detection accuracy of existing CD methods significantly declines. 
Furthermore, our analysis shows that noisy edges not only degrade performance but also cause originally normal edges to behave like noisy ones, leading their incident nodes to drift away from their true community structures.
We refer to these as \textit{misaligned edges}, which further amplify distortions in the overall topology (more details in Section~\ref{s5}.2).

\vspace{1mm}
\noindent \textbf{Proposed Framework.}
To address this limitation, we propose {\rec} (Community Refinement and Contrastive Learning), a model-agnostic post-processing framework designed to refine and enhance the community structures produced by any CD method on signed networks.
Once the initial community structure obtained from a target CD method is provided to \textsf{ReCon}, it progressively refines the structure through four iterative steps: 
(1) \textit{structural refinement}, which reassigns node memberships from both local and global perspectives to enhance structural balance;
(2) \textit{boundary refinement}, which identifies nodes that violate community consistency and relocates them based on balance theory;
(3) \textit{contrastive learning}, which applies multi-view self-supervised alignment to both node- and community-level embeddings to refine representation quality and strengthen community separation; and
(4) \textit{clustering}, which derives the final community structure from the learned embeddings.
% Through this progressive refinement, \textsf{ReCon} effectively reorganizes noisy and inconsistent communities into more coherent and well-separated structures.

\vspace{1mm}
\noindent \textbf{Contributions.}
The contributions of this paper are as follows:
% \vspace{-1mm}
\begin{itemize}[leftmargin=*]
    \item \textbf{Important Observation.} Existing CD methods are vulnerable to noisy edges in signed networks, which cause edges to become misaligned with community structures, ultimately degrading overall detection accuracy.
    \item \textbf{Novel Framework.} We propose \textsf{ReCon}, a model-agnostic framework that enhances the quality of CD results produced by any algorithm on signed networks through community refinement and contrastive learning.
    \item \textbf{Extensive Evaluation.} We conduct comprehensive experiments on 18 synthetic and 4 real-world signed networks using 4 CD methods, and the results demonstrate that \textsf{ReCon} consistently improves all performance metrics across diverse CD methods.
\end{itemize}
% To ensure full reproducibility, we publicly release the source code together with all parameter settings and implementation details at: https://buly.kr/BTQZDBY.

\vspace{-3mm}

%% file: s2.tex
\section{Related Work}~\label{s2}
In signed networks, community detection (CD) partitions nodes into groups with predominantly positive intra-community and negative inter-community edges~\cite{yan17:tkde, kan23:tkde}. 
FEC~\cite{yan17:tkde} solves the \textit{k}-way partition via the signed Laplacian; 
SPONGE~\cite{coc19:aistats} relaxes it into a generalized eigenproblem;
SSSNET~\cite{he22:sdm} jointly optimizes semi-supervised representation learning and clustering; 
and DSGC~\cite{Zha25:www} integrates weak balance theory with noise-aware edge refinement. 
Despite their effectiveness, these methods remain sensitive to noisy edges since all edges directly participate in the CD process.

\vspace{-2mm}

%% file: s4.tex
\section{\textsf{ReCon}: Proposed Framework}~\label{s4}
\begin{figure}[t]
\centering
\includegraphics[width=0.94\columnwidth]{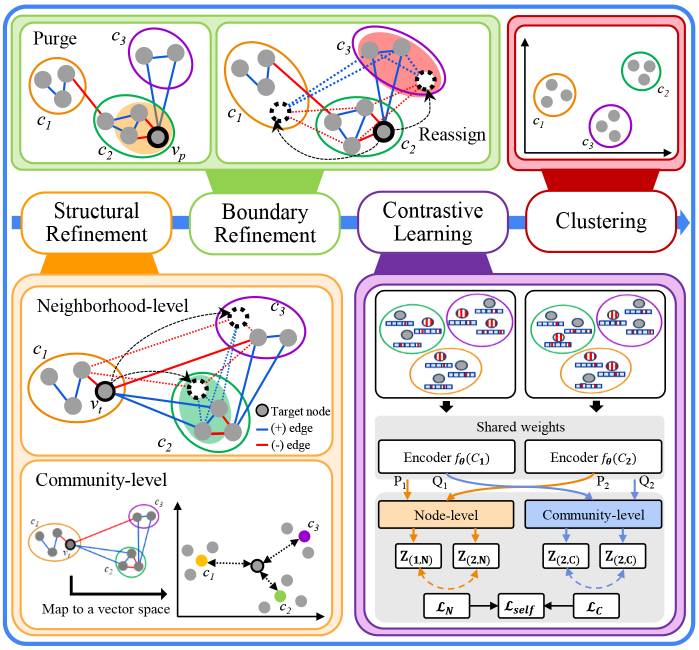}
\vspace{-4mm}
\caption{Overview of {\rec}.} \label{fig_1}
%\label{f1}
\vspace{-7mm}
% \vspace{0.4cm}
\end{figure}

\noindent \textbf{Overview.} In this section, we present {\rec}, a post-processing framework designed to refine community structures by detecting and reassigning misaligned edges.
Figure~\ref{fig_1} illustrates the overall process in {\rec}, consisting of four main steps: \textbf{(S1)} structural refinement, \textbf{(S2)} boundary refinement, \textbf{(S3)} contrastive learning, and \textbf{(S4)} clustering.
Once the community structure obtained from the target CD method is input to {\rec}, it performs these four steps iteratively to refine the node vectors in a progressive way.

\vspace{1mm}
\noindent \textbf{Structural Refinement.}
\textit{Structural refinement} reassigns node memberships from two complementary perspectives: \textit{neighborhood-level} (\ie, local) and \textit{community-level} (\ie, global).
At the neighborhood-level, for a target node $v_i$, {\rec} computes an \textit{N-Score} for each community based on the signs of its adjacent edges. 
The N-score is defined as:
\vspace{-2mm}
\begin{equation}
\text{N-Score}(v_i, C_k)
= \frac{1}{d_{v_i}}
\sum_{v_j \in \mathcal{N}(v_i)} s_{ij} \cdot \delta(e_{i,j}, C_k),
\end{equation}
where $\mathcal{N}(v_i)$ denotes the set of all neighbors of node $v_i$, 
$s_{ij} \in \{+1, -1\}$ represents the edge sign, 
$\delta(e_{i,j}, C_k)$ is the Kronecker delta function that equals $1$ if $e_{i,j} \in C_k$ and $0$ otherwise; 
intuitively, it indicates whether the edge between nodes $v_i$ and $v_j$ is considered to lie within community $C_k$, 
and $d_{v_i}$ denotes the degree of $v_i$.
The node is likely to be reassigned to the community with the highest N-score, thereby maximizing local structural balance.

At the community level, {\rec} computes a \textit{C-Score} between each node’s embedding and every community centroid, and reassigns the node to the community with the highest similarity.
The C-Score is defined as:
\begin{equation}
\text{C-Score}(v_i, C_k)
= \mathbf{c}_k^{\top} \mathbf{z}_{v_i},
\end{equation}
where $\mathbf{c}_k$ denotes the centroid vector of community $C_k$ 
and $\mathbf{z}_{v_i}$ is the embedding vector of node $v_i$. 
The node is likely to be reassigned to the community with the highest C-Score, thereby maximizing global representational balance and aligning node embeddings with community-level semantics. 

Finally, {\rec} combines the two scores via a weighted summation and assigns the node to the community with the highest combined score.
Subsequently, a \textit{softmax function} is applied to convert these combined scores into a probability distribution, and the node is reassigned to the community with the highest probability.
This process reorganizes the overall structure toward greater local–global consistency.

\vspace{1mm}
\noindent \textbf{Boundary Refinement.}
Since nodes are reassigned based on probability, some nodes may still violate community consistency after structural refinement, remaining near the community boundaries (\eg, node $v_p$ in Figure~\ref{fig_1}).
\textit{Boundary refinement} reassigns these nodes to reinforce boundary integrity. 
To detect nodes requiring adjustment, {\rec} leverages balanced triangles: if three nodes form a (+––)-triangle within the same community, the node connected only by negative edges is likely misassigned and becomes a reassignment target.
When balance theory is inapplicable (\eg, nodes not in any triangle), {\rec} computes each node’s purge likelihood based on the proportion of edges that violate community characteristics.
Nodes identified by these two criteria are then selected as purge candidates and reassigned to the community that maximizes the number of balanced (+++)-triangles within that community, thereby enhancing overall structural balance.

\vspace{1mm}
\noindent \textbf{Contrastive Learning.}
While structural and boundary refinement improve discrete assignments, \textit{contrastive learning} further reshapes the embedding space to amplify community separability and reduce representation-level noise.
To this end, {\rec} generates two augmented views of the refined community structure using stochastic functions that apply random node-feature and community masking. 
An encoder then produces node and community embeddings through a two-stage aggregation capturing intra-community and inter-community semantics. 

{\rec} aligns representations across views at both the \textit{node} and \textit{community} levels. 
For node-level contrast, the embedding of node $v_i$ from the first view, $\mathbf{z}_{1,i}$, is the anchor, its counterpart $\mathbf{z}_{2,i}$ the positive, and all other embeddings $\mathbf{z}_{2,k}$ $(k\!\ne\! i)$ negatives.
The loss for node $v_i$ is defined as:
\begin{equation}
\ell_n(\mathbf{z}_{1,i},\mathbf{z}_{2,i})
=-\log
\frac{\exp(s(\mathbf{z}_{1,i},\mathbf{z}_{2,i})/\tau_n)}
{\sum_{k=1}^{|V|}\exp(s(\mathbf{z}_{1,i},\mathbf{z}_{2,k})/\tau_n)},
\end{equation}
where $\tau_n$ is the temperature and $s(\cdot,\cdot)$ the cosine similarity.  
The overall node-level loss averages over all nodes and both (potentially asymmetric) views:
\begin{equation}\label{ln}
\mathcal{L}_n=\frac{1}{2|V|}
\sum_{i=1}^{|V|}
\big[\ell_n(\mathbf{z}_{1,i},\mathbf{z}_{2,i})
+\ell_n(\mathbf{z}_{2,i},\mathbf{z}_{1,i})\big].
\end{equation}

Similarly, at the community level, the embedding $\mathbf{y}_{1,j}$ from the first view is the anchor, $\mathbf{y}_{2,j}$ the positive, and all $\mathbf{y}_{2,k}$ $(k\!\ne\! j)$ negatives. The loss for community $c_j$ is defined as:
\begin{equation}
\ell_c(\mathbf{y}_{1,j},\mathbf{y}_{2,j})
=-\log
\frac{\exp(s(\mathbf{y}_{1,j},\mathbf{y}_{2,j})/\tau_c)}
{\sum_{k=1}^{|C|}\exp(s(\mathbf{y}_{1,j},\mathbf{y}_{2,k})/\tau_c)},
\end{equation}
where $\tau_c$ is the temperature.  
The overall community-level loss is
\begin{equation}\label{lc}
\mathcal{L}_c=\frac{1}{2|C|}
\sum_{j=1}^{|C|}
\big[\ell_c(\mathbf{y}_{1,j},\mathbf{y}_{2,j})
+\ell_c(\mathbf{y}_{2,j},\mathbf{y}_{1,j})\big].
\end{equation}
The total contrastive loss is defined as:
\begin{equation}\label{final_lh}
\mathcal{L}_{cl}=\omega_n\mathcal{L}_n+\omega_c\mathcal{L}_c.
\end{equation}

\vspace{1mm}
\noindent \textbf{Clustering.}
\textit{Clustering} is a step that derives the final community structure by applying a clustering algorithm to the node representations obtained from the contrastive learning stage. 
Specifically, we employ the \textit{k}-means algorithm on the learned embeddings to produce the final community assignments.

\vspace{-3mm}

%% file: s5.tex
\section{Evaluation}~\label{s5}
We evaluate {\rec} by answering the following questions:

\begin{itemize}[leftmargin=*]
    \item \textbf{EQ1}: Do existing CD methods really suffer from noisy edges?
    \item \textbf{EQ2}: Are the steps \textbf{(S1)}, \textbf{(S2)}, and \textbf{(S3)} in {\rec} effective?
    \item \textbf{EQ3}: To what extent does {\rec} improve the accuracy of CD across different algorithms and graph properties?
\end{itemize}

\vspace{-2mm}
\subsection{Experimental Settings}\label{4.1}
\noindent \textbf{Datasets and CD Methods.}
Following~\cite{coc19:aistats, he22:sdm, Zha25:www}, we evaluate \textsf{ReCon} on 18 synthetic and 4 real-world signed networks.
Synthetic networks are generated via the \textit{Signed Stochastic Block Model} (SSBM)~\cite{coc19:aistats}, where we vary the number of nodes ($|V|$), communities ($|C|$), edge density ($p$), and noise ratio ($\mu$), adopting the same settings as~\cite{Zha25:www} for fair comparison.
For real-world datasets, we use Rainfall, BitcoinOTC, PPI, and Wiki-RfA~\cite{he22:sdm, ku16:icdm}.
As baseline CD methods, we employ FEC~\cite{yan17:tkde}, SPONGE~\cite{coc19:aistats}, SSSNET~\cite{he22:sdm}, and DSGC~\cite{Zha25:www}.
For CD methods, we use the source codes provided by the authors of the original papers~\cite{coc19:aistats, he22:sdm, Zha25:www}.
All hyperparameters are configured using the best settings identified through an extensive grid search within the ranges recommended in the corresponding papers.
To ensure full reproducibility, we publicly release the source code together with all parameter settings and implementation details at: https://buly.kr/BTQZDBY.

\vspace{1mm}
\noindent \textbf{Evaluation Protocol.} For each signed network, we run each CD baseline to obtain the initial community structure and apply {\rec} to obtain the refined one.
For synthetic networks with ground-truth, we evaluate accuracy using \textit{ARI}; for real-world networks without ground-truth, we use \textit{modularity}.
% For \textit{k}-means, we set \textit{k} to the number of ground-truth communities for synthetic networks where ground truth is available.
% For real-world networks, we follow the same selection strategy as in prior studies~\cite{coc19:aistats, he22:sdm, Zha25:www} to determine the value of \textit{k}.
All results are reported as the mean and standard deviation over five independent runs.

\vspace{-2mm}
\subsection{Experimental Results}\label{4.1}
\noindent \textbf{EQ1: Impact of Noisy Edges on CD.}
To examine the impact of noisy edges on CD, we generate synthetic signed networks using SSBM~\cite{coc19:aistats}, where the proportion of noisy edges is systematically controlled to assess robustness.
Following~\cite{coc19:aistats, Zha25:www}, we fix $|V|=1{,}000$, $|C|=5$, and $p=0.01$, while varying $\mu$ from 0 to 0.2 in increments of 0.01.
Then, we conduct four CD methods to the generated networks and measure the accuracy.
Figure~\ref{tab_2}-(a) shows that the accuracy of all methods consistently decreases as the noise ratio increases.
To gain further insight, we analyze the detected community structures by measuring the proportion of \textit{misaligned edges}.
As shown in Figure~\ref{tab_2}-(b), higher noise levels result in a larger fraction of misaligned edges due to cascading noise effects.
% Notably, in SPONGE, 70–80\% of the misaligned edges across all noise levels are normal rather than truly noisy—a tendency also observed in other CD methods. 
This finding indicates that noisy edges primarily degrade CD accuracy by distorting the community structure, which in turn causes normal edges to appear misaligned with incorrect community assignments.

\input{Figs/f2}
\begin{table}[t]
\centering
\caption{The ablation study of {\rec} (ARI $\times$ 100)}
\label{tab_1}
\scriptsize
\vspace{-3mm}
\def\arraystretch{1.25}
\resizebox{\columnwidth}{!}{
\begin{tabular}{ccc||cccc|c}
\toprule
 SR & BR & CL & \textbf{FEC} & \textbf{SPONGE} & \textbf{SSSNET} & \textbf{DSGC} & \textbf{AR}  \\ 
\midrule
\midrule
\textcolor{red}{\xmark}  & \textcolor{red}{\xmark} & \textcolor{red}{\xmark} 
&  30.31 $\pm$ 1.25 (7) &34.18 $\pm$ 1.69 (7) & 48.03 $\pm$ 2.91 (8) & 62.15 $\pm$ 1.28 (8) & 7.5   \\
\textcolor{teal}{\checkmark}  & \textcolor{red}{\xmark} & \textcolor{red}{\xmark}
&  36.40 $\pm$ 1.54 (6) & 45.56 $\pm$ 5.49 (4) & 60.65 $\pm$ 4.57 (4) & 69.86 $\pm$ 3.00 (4) & 4.5   \\
\textcolor{red}{\xmark} & \textcolor{teal}{\checkmark}  & \textcolor{red}{\xmark}
&  36.95 $\pm$ 2.15 (4) & 40.12 $\pm$ 1.00 (5) & 55.77 $\pm$ 3.67 (5) & 69.47 $\pm$ 0.98 (5) & 4.8   \\
\textcolor{red}{\xmark} & \textcolor{red}{\xmark} & \textcolor{teal}{\checkmark} 
&  30.31 $\pm$ 1.25 (7) & 34.18 $\pm$ 1.69 (7) & 47.96 $\pm$ 2.93 (7) & 63.31 $\pm$ 0.55 (7) & 7.0   \\
\textcolor{teal}{\checkmark} & \textcolor{teal}{\checkmark} & \textcolor{red}{\xmark}
&  45.92 $\pm$ 1.10 (2) & 52.90 $\pm$ 5.90 (2) & 61.02 $\pm$ 4.66 (2) & 73.92 $\pm$ 1.28 (3) & 2.3   \\
\textcolor{teal}{\checkmark} & \textcolor{red}{\xmark} & \textcolor{teal}{\checkmark}
&  38.85 $\pm$ 1.39 (3) & 46.01 $\pm$ 5.94 (3) & 63.91 $\pm$ 2.04 (2) & 70.32 $\pm$ 4.33 (3) & 2.8   \\
\textcolor{red}{\xmark} & \textcolor{teal}{\checkmark} & \textcolor{teal}{\checkmark} 
&  36.50 $\pm$ 2.14 (5) & 39.98 $\pm$ 1.62 (6) & 54.24 $\pm$ 4.06 (6) & 67.45 $\pm$ 1.55 (6) & 5.8   \\
\textcolor{teal}{\checkmark} & \textcolor{teal}{\checkmark} & \textcolor{teal}{\checkmark} 
&  \textbf{48.89 $\pm$ 3.11} (1) & \textbf{55.53 $\pm$ 4.87} (1) & \textbf{67.75 $\pm$ 3.05} (1) & \textbf{73.57 $\pm$ 2.68} (1) &\textbf{1.0}   \\
\bottomrule
\end{tabular}
}
\vspace{-6mm}
\end{table}

\begin{table*}[t]
\centering
\caption{Community detection accuracy (ARI $\times$ 100) in synthetic signed networks}
\label{tab_2}
\footnotesize
\vspace{-4mm}
\def\arraystretch{1.1}
\resizebox{\textwidth}{!}{
  \input{Tabs/t2.tex}
}
\vspace{-0.4cm}
\end{table*}

\begin{table*}[t]
\centering
\caption{Community detection quality (modularity) in real-world signed networks}
\label{tab_3}
\footnotesize
\vspace{-4mm}
\def\arraystretch{1.1}
\resizebox{\textwidth}{!}{
  \input{Tabs/t3.tex}
}
\vspace{-0.5cm}
\end{table*}

\vspace{1mm}
\noindent \textbf{EQ2: Ablation Study.}
In this experiment, we evaluate the effectiveness of each key-step incorporated in {\rec}.
Following~\cite{coc19:aistats, he22:sdm, Zha25:www}, we use a synthetic signed network generated by SSBM with $|V|=1{,}000$, $|C|=5$, $p=0.01$, and $\mu=0.02$.
% \footnote{Due to space limitations, we report results only on a baseline synthetic network, but similar trends are observed across other synthetic and real-world datasets.}
Table~\ref{tab_1} summarizes the ablation results.
First, {\rec} with all three steps yields the highest accuracy across all datasets and CD methods.
Second, using any single step generally outperforms the case with none, while applying only contrastive learning (CL) may even degrade performance.
This degradation occurs because CL operates on misdetected communities, drawing incorrectly grouped nodes closer and amplifying structural errors.
These findings underscore the importance of re-adjusting misaligned edges for accurate boundary recovery.
Third, combining two steps (\eg, SR+BR or SR+CL) further improves accuracy, whereas BR+CL performs worse than either alone, since the absence of SR leaves major structural errors unresolved—causing BR to fine-tune flawed structures and CL to reinforce them.

\vspace{1mm}
\noindent \textbf{EQ3: Accuracy Comparison.}
Finally, we evaluate how {\rec} enhances community detection accuracy across different CD methods and network properties.
Table~\ref{tab_2} summarizes the results on 18 \textit{synthetic} signed networks, showing that the refined community structure $C_{R}$ obtained with {\rec} yields substantial performance gains over the original structure $C$ in almost all cases (70 out of 72 network–CD combinations).
Table~\ref{tab_3} reports the modularity scores on four \textit{real-world} signed networks, where incorporating {\rec} yields improvements across all network–CD method combinations.
Figure~\ref{fig_3} visualizes the CD outcomes of FEC and SPONGE on Rainfall.
Integrating {\rec} (Figures~\ref{fig_3}-(b) and (d)) exhibits much clearer and more compact community separations than their baselines (Figures~\ref{fig_3}-(a) and (c)), confirming that our contrastive learning step promotes more distinct and accurate community representations.

\vspace{-6mm}

%% file: Figs/f2.tex
\begin{figure}[t]
\centering

% ==== 공유 legend (위쪽) ====
\pgfplotslegendfromname{sharedlegend}
\vspace{-0.2cm}

\begin{tabular}{cc}
% ---- (왼쪽) ----
\begin{tikzpicture}
\scriptsize
\begin{axis}[
    width=4.2cm,height=2.4cm,
    ymin=0, ymax=100, xmin=0, xmax=0.20,
    ylabel={ARI $\times$ 100},
    xlabel={Noise rate (\%)},
    legend to name=sharedlegend,           % << 공유 legend로 내보내기
    legend style={
        font=\scriptsize, draw=none, fill=none,
        legend columns=4,
        /tikz/every even column/.append style={column sep=0.25cm}
    },
]
    \addplot[color=black, line width=0.8pt, mark=square*, mark size=1pt, smooth] file {fec.txt}; 
    \addplot[color=blue,  line width=0.8pt, mark=*,        mark size=1pt, smooth] file {sponge.txt};
    \addplot[color=orange,line width=0.8pt, mark=pentagon*, mark size=1pt, smooth] file {sssnet.txt};
    \addplot[color=red,   line width=0.8pt, mark=triangle*, mark size=1pt, smooth] file {dsgc.txt};
    \legend{\textit{FEC}, \textit{SPONGE}, \textit{SSSNET}, \textit{DSGC}}
\end{axis}
\end{tikzpicture}
\hspace{0.4cm}
&
% ---- (오른쪽) ----
\begin{tikzpicture}
\scriptsize
\begin{axis}[
    width=4.2cm,height=2.4cm,
    ymin=0, ymax=30, xmin=0, xmax=0.20,
    ylabel={M. edge ratio (\%)},
    xlabel={Noise rate (\%)},
]
    \addplot[color=black, line width=0.8pt, mark=square*,  mark size=1pt, smooth] file {fecm.txt}; 
    \addplot[color=blue,  line width=0.8pt, mark=*,         mark size=1pt, smooth] file {spongem.txt};
    \addplot[color=orange,line width=0.8pt, mark=pentagon*,  mark size=1pt, smooth] file {sssnetm.txt};
    \addplot[color=red,   line width=0.8pt, mark=triangle*,  mark size=1pt, smooth] file {dsgcm.txt};
\end{axis}
\end{tikzpicture}
\\
\textbf{(a) \footnotesize{Accuarcy}} & \textbf{(b) \footnotesize{Misaligned edge ratio}}
\end{tabular}

\vspace{-0.4cm}
\caption{CD accuracy and misaligned edge rates under increasing noise rates.}
\label{fig_2}
\vspace{-0.4cm}
\end{figure}
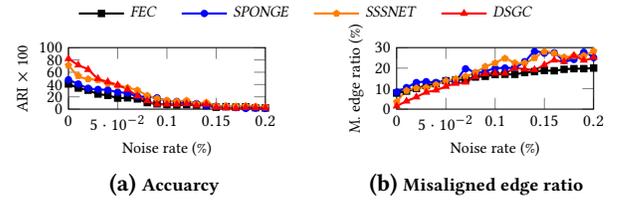

%% file: Tabs/t2.tex
\begin{tabular}{cc|ccc|ccc|ccc|ccc}
\toprule
 &  & \multicolumn{3}{c|}{\textbf{FEC}} & \multicolumn{3}{c|}{\textbf{SPONGE}} & \multicolumn{3}{c|}{\textbf{SSSNET}} & \multicolumn{3}{c}{\textbf{DSGC}} \\
 && $C$  & $C_R$ & Gain (\%)  & $C$  & $C_R$  & Gain (\%) & $C$  & $C_R$ & Gain (\%) & $C$  & $C_R$ & Gain (\%) \\
\midrule\midrule
\multirow{5}{*}{\textbf{{$|V|$}}}
&SSBM-300-5-0.01-0.00 & 4.20 $\pm$ 1.51 & 5.53 $\pm$ 2.03 & \textcolor{blue}{31.67} & 0.42 $\pm$ 0.68 & 5.08 $\pm$ 0.45 & \textcolor{blue}{1109.52} & 9.97 $\pm$ 3.37 & 11.49 $\pm$ 2.56 & \textcolor{blue}{15.25} & 8.48 $\pm$ 3.48 & 8.66 $\pm$ 2.94 & \textcolor{blue}{2.12} \\
&SSBM-500-5-0.01-0.00 & 10.37 $\pm$ 1.85 & 13.98 $\pm$ 1.42 & \textcolor{blue}{34.81} & 1.68 $\pm$ 3.60 & 9.87 $\pm$ 7.00 & \textcolor{blue}{487.50} & 21.90 $\pm$ 3.17 & 24.50 $\pm$ 3.67 & \textcolor{blue}{11.87} & 26.81 $\pm$ 2.17 & 26.91 $\pm$ 3.32 & \textcolor{blue}{0.37} \\
&SSBM-800-5-0.01-0.00 & 19.31 $\pm$ 2.58 & 31.96 $\pm$ 4.45 & \textcolor{blue}{65.51} & 29.88 $\pm$ 3.65 & 49.39 $\pm$ 5.54 & \textcolor{blue}{65.29} & 42.54 $\pm$ 3.18 & 63.56 $\pm$ 4.27 & \textcolor{blue}{49.41} & 62.79 $\pm$ 2.27 & 74.39 $\pm$ 1.34 & \textcolor{blue}{18.47} \\
&SSBM-1000-5-0.01-0.00 & 42.24 $\pm$ 2.00 & 72.93 $\pm$ 8.20 & \textcolor{blue}{72.66} & 48.18 $\pm$ 2.53 & 73.30 $\pm$ 5.82 & \textcolor{blue}{52.14} & 70.12 $\pm$ 3.78 & 89.30 $\pm$ 1.02 & \textcolor{blue}{27.35} & 81.26 $\pm$ 0.52 & 90.18 $\pm$ 0.99 & \textcolor{blue}{10.98} \\
&SSBM-1200-5-0.01-0.00 & 52.53 $\pm$ 2.62 & 76.70 $\pm$ 10.46 & \textcolor{blue}{46.01} & 67.26 $\pm$ 2.24 & 84.29 $\pm$ 3.62 & \textcolor{blue}{25.32} & 76.63 $\pm$ 3.97 & 89.82 $\pm$ 3.79 & \textcolor{blue}{17.21} & 87.47 $\pm$ 0.50 & 90.99 $\pm$ 1.52 & \textcolor{blue}{4.02} \\
\midrule
\multirow{5}{*}{\textbf{{$|C|$}}}
&SSBM-1000-4-0.01-0.02 & 78.95 $\pm$ 0.28 & 94.10 $\pm$ 0.27 & \textcolor{blue}{19.19} & 84.14 $\pm$ 0.20 & 93.87 $\pm$ 0.37 & \textcolor{blue}{11.56} & 87.43 $\pm$ 1.15 & 95.03 $\pm$ 0.86 & \textcolor{blue}{8.69} & 85.06 $\pm$ 0.26 & 94.97 $\pm$ 0.54 & \textcolor{blue}{11.65} \\
&SSBM-1000-5-0.01-0.02 &  30.31 $\pm$ 1.25 & 48.89 $\pm$ 3.11 & \textcolor{blue}{61.30} & 34.18 $\pm$ 1.69 & 55.53 $\pm$ 4.87 & \textcolor{blue}{62.46} & 48.03 $\pm$ 2.91 & 67.75 $\pm$ 3.05 & \textcolor{blue}{41.06} & 62.15 $\pm$ 1.28 & 73.57 $\pm$ 2.68 & \textcolor{blue}{18.37} \\
&SSBM-1000-6-0.01-0.02 & 12.96 $\pm$ 3.18 & 19.92 $\pm$ 6.75 & \textcolor{blue}{53.70} & 14.31 $\pm$ 1.32 & 23.95 $\pm$ 4.19 & \textcolor{blue}{67.37} & 28.38 $\pm$ 2.70 & 36.75 $\pm$ 1.07 & \textcolor{blue}{29.49} & 45.19 $\pm$ 2.75 & 48.03 $\pm$ 2.57 & \textcolor{blue}{6.28} \\
&SSBM-1000-7-0.01-0.02 & 4.45 $\pm$ 0.48 & 6.87 $\pm$ 2.26 & \textcolor{blue}{54.38} & 7.03 $\pm$ 1.17 & 13.16 $\pm$ 1.84 & \textcolor{blue}{87.20} & 16.86 $\pm$ 4.53 & 22.02 $\pm$ 5.06 & \textcolor{blue}{30.60} & 19.15 $\pm$ 3.06 & 17.87 $\pm$ 4.83 & \textcolor{red}{-6.68} \\
&SSBM-1000-8-0.01-0.02 & 2.76 $\pm$ 0.22 & 4.27 $\pm$ 0.44 & \textcolor{blue}{54.71} & 5.75 $\pm$ 0.65 & 6.97 $\pm$ 0.91 & \textcolor{blue}{21.22} & 11.44 $\pm$ 1.06 & 12.06 $\pm$ 2.02 & \textcolor{blue}{5.42} & 10.38 $\pm$ 1.48 & 8.38 $\pm$ 1.21 & \textcolor{red}{-19.27} \\
\midrule
\multirow{5}{*}{{\textbf{{$p$}}}}
&SSBM-1000-10-0.01-0.02 & 1.66 $\pm$ 0.57 & 2.90 $\pm$ 0.61 & \textcolor{blue}{74.70} & 1.35 $\pm$ 0.35 & 3.22 $\pm$ 1.13 & \textcolor{blue}{138.52} & 9.01 $\pm$ 1.48 & 9.09 $\pm$ 2.77 & \textcolor{blue}{0.89} & 6.08 $\pm$ 1.48 & 6.18 $\pm$ 2.01 & \textcolor{blue}{1.64} \\
&SSBM-1000-10-0.02-0.02 & 7.88 $\pm$ 1.20 & 15.01 $\pm$ 3.81 & \textcolor{blue}{90.48} & 10.19 $\pm$ 1.23 & 17.26 $\pm$ 2.88 & \textcolor{blue}{69.38} & 26.53 $\pm$ 2.15 & 31.43 $\pm$ 2.01 & \textcolor{blue}{18.47} & 45.37 $\pm$ 1.84 & 45.86 $\pm$ 3.22 & \textcolor{blue}{1.08} \\
&SSBM-1000-10-0.03-0.02 & 25.86 $\pm$ 3.47 & 52.56 $\pm$ 7.00 & \textcolor{blue}{103.25} & 41.39 $\pm$ 2.52 & 68.46 $\pm$ 8.12 & \textcolor{blue}{65.40} & 66.18 $\pm$ 1.74 & 83.50 $\pm$ 3.79 & \textcolor{blue}{26.17} & 77.34 $\pm$ 0.30 & 86.72 $\pm$ 5.22 & \textcolor{blue}{12.13} \\
&SSBM-1000-10-0.04-0.02 & 46.14 $\pm$ 1.68 & 90.36 $\pm$ 5.82 & \textcolor{blue}{95.84} & 60.95 $\pm$ 6.51 & 89.24 $\pm$ 5.26 & \textcolor{blue}{46.42} & 85.61 $\pm$ 1.47 & 94.75 $\pm$ 0.46 & \textcolor{blue}{10.68} & 89.08 $\pm$ 0.38 & 94.12 $\pm$ 0.17 & \textcolor{blue}{5.66} \\
&SSBM-1000-10-0.05-0.02 & 71.02 $\pm$ 6.87 & 84.18 $\pm$ 9.77 & \textcolor{blue}{18.53} & 86.95 $\pm$ 12.05 & 94.70 $\pm$ 11.38 & \textcolor{blue}{8.91} & 96.82 $\pm$ 1.54 & 99.68 $\pm$ 0.27 & \textcolor{blue}{2.95} & 95.72 $\pm$ 3.45 & 97.60 $\pm$ 4.70 & \textcolor{blue}{1.96} \\
\midrule
\multirow{3}{*}{{\textbf{{$\mu$}}}}
&SSBM-1000-5-0.01-0.04 & 23.94 $\pm$ 1.19 & 35.58 $\pm$ 2.24 & \textcolor{blue}{48.62} & 31.61 $\pm$ 2.83 & 50.01 $\pm$ 3.86 & \textcolor{blue}{58.21} & 40.79 $\pm$ 4.11 & 56.99 $\pm$ 2.52 & \textcolor{blue}{39.72} & 43.79 $\pm$ 0.58 & 55.80 $\pm$ 4.23 & \textcolor{blue}{27.43} \\
&SSBM-1000-5-0.01-0.06 & 18.44 $\pm$ 3.04 & 24.98 $\pm$ 6.16 & \textcolor{blue}{35.47} & 25.96 $\pm$ 1.23 & 35.43 $\pm$ 2.34 & \textcolor{blue}{36.48} & 34.24 $\pm$ 1.90 & 46.86 $\pm$ 2.60 & \textcolor{blue}{36.86} & 32.22 $\pm$ 1.28 & 40.87 $\pm$ 3.30 & \textcolor{blue}{26.85} \\
&SSBM-1000-5-0.01-0.08 & 11.42 $\pm$ 1.54 & 13.50 $\pm$ 2.62 & \textcolor{blue}{18.21} & 18.86 $\pm$ 1.68 & 29.91 $\pm$ 2.33 & \textcolor{blue}{58.59} & 22.99 $\pm$ 1.56 & 29.44 $\pm$ 1.99 & \textcolor{blue}{28.06} & 12.78 $\pm$ 1.09 & 16.79 $\pm$ 2.70 & \textcolor{blue}{31.38} \\

\midrule
\bottomrule

\end{tabular}%

%% file: Tabs/t3.tex
\begin{tabular}{c|ccc|ccc|ccc|ccc}
\toprule
 & \multicolumn{3}{c|}{\textbf{FEC}} & \multicolumn{3}{c|}{\textbf{SPONGE}} & \multicolumn{3}{c|}{\textbf{SSSNET}} & \multicolumn{3}{c}{\textbf{DSGC}} \\
 & $C$ & $C_R$ & Gain (\%) & $C$ & $C_R$ & Gain (\%) & $C$ & $C_R$ & Gain (\%) & $C$ & $C_R$ & Gain (\%) \\
\midrule\midrule
Rainfall & 0.5454 $\pm$ 0.0235 & 0.6734 $\pm$ 0.0227 & \textcolor{blue}{23.46}  & 0.5159 $\pm$ 0.0000 & 0.6690 $\pm$ 0.0147  & \textcolor{blue}{29.67} & 0.6168 $\pm$ 0.0417 & 0.6752 $\pm$ 0.0200  &  \textcolor{blue}{9.47} & 0.5048 $\pm$ 0.0685 & 0.6582 $\pm$ 0.0610 &\textcolor{blue}{30.39}\\
BitcoinOTC & 0.0341 $\pm$ 0.0402  & 0.1835 $\pm$ 0.1364 & \textcolor{blue}{438.12}  & 0.0846 $\pm$ 0.0208 & 0.0904 $\pm$ 0.0755  & \textcolor{blue}{6.86} & 0.1535 $\pm$ 0.0286 &  0.2136 $\pm$ 0.0329 &  \textcolor{blue}{39.15} & 0.0683 $\pm$ 0.0052 & 0.0975 $\pm$ 0.0814 &\textcolor{blue}{42.75}\\ 
PPI & 0.0219 $\pm$ 0.0056  & 0.1506 $\pm$ 0.1007 & \textcolor{blue}{587.67}  & 0.0338 $\pm$ 0.0033 & 0.1516 $\pm$ 0.1609  & \textcolor{blue}{348.52} & 0.3153 $\pm$ 0.0142 &  0.3909 $\pm$ 0.1756 &  \textcolor{blue}{23.98} & 0.3369 $\pm$ 0.0527 & 0.4314 $\pm$ 0.2507 &\textcolor{blue}{35.66}\\
Wiki-Rfa & 0.0279 $\pm$ 0.0003  & 0.0690 $\pm$ 0.0532 & \textcolor{blue}{2325.81}  & 0.0020 $\pm$ 0.0014 & 0.0848 $\pm$ 0.0670  & \textcolor{blue}{4140.00} & 0.0481 $\pm$ 0.0000 &  0.0515 $\pm$ 0.0307 &  \textcolor{blue}{7.07} &0.0180 $\pm$ 0.0032 & 0.0341 $\pm$ 0.0046 &\textcolor{blue}{89.44}\\
\midrule
\bottomrule

\end{tabular}%

%% file: s6.tex
\section{Conclusions}~\label{s6}
In this paper, we showed that existing CD methods for signed networks are highly vulnerable to noisy edges, which generate misaligned edges that distort community structures. To address this issue, we proposed \textsf{ReCon}, a post-processing framework that detects and reassigns misaligned edges to restore community consistency. {\rec} adaptively refines the initial community assignments and applies multi-view contrastive learning to obtain more discriminative representations. Extensive experiments validate that \textsf{ReCon} is a simple yet highly effective plug-in module that consistently improves accuracy across diverse signed CD methods and network conditions, without requiring any algorithm-specific modifications.

\vspace{-3mm}

\input{Figs/f3}

% \newpage
% \input{Link/t10}
% \input{Link/t11}
% \input{Link/t12}

%% file: Figs/f3.tex
\begin{figure}[t]
\centering
\scriptsize
\begin{subfigure}[t]{0.22\textwidth}
    \centering
    \includegraphics[width=3.5cm,height=1.1cm]{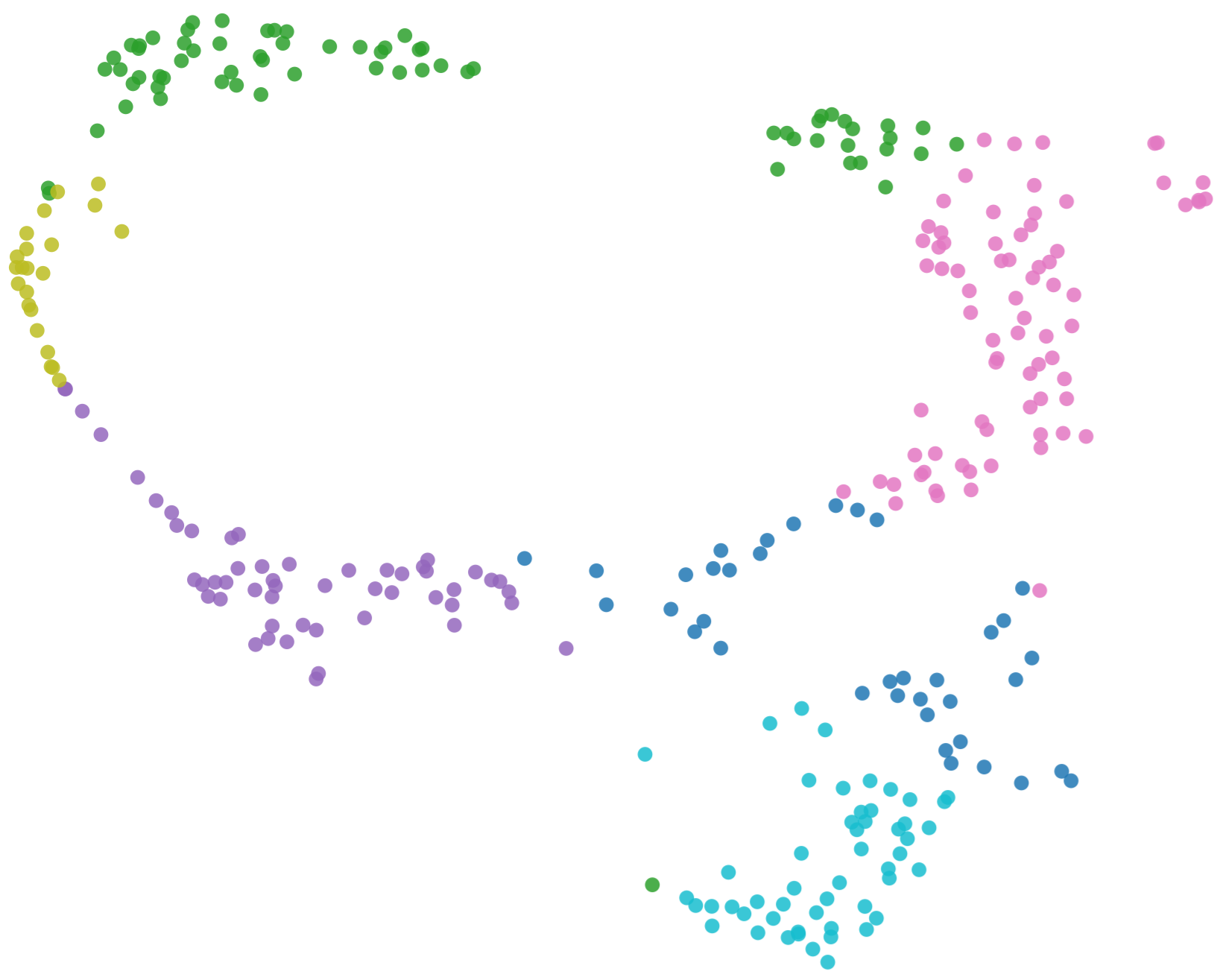}
    \caption{\scriptsize{FEC}}
    \label{f3c}
\end{subfigure}
\hfill
\begin{subfigure}[t]{0.22\textwidth}
    \centering
    \includegraphics[width=3.5cm,height=1.1cm]{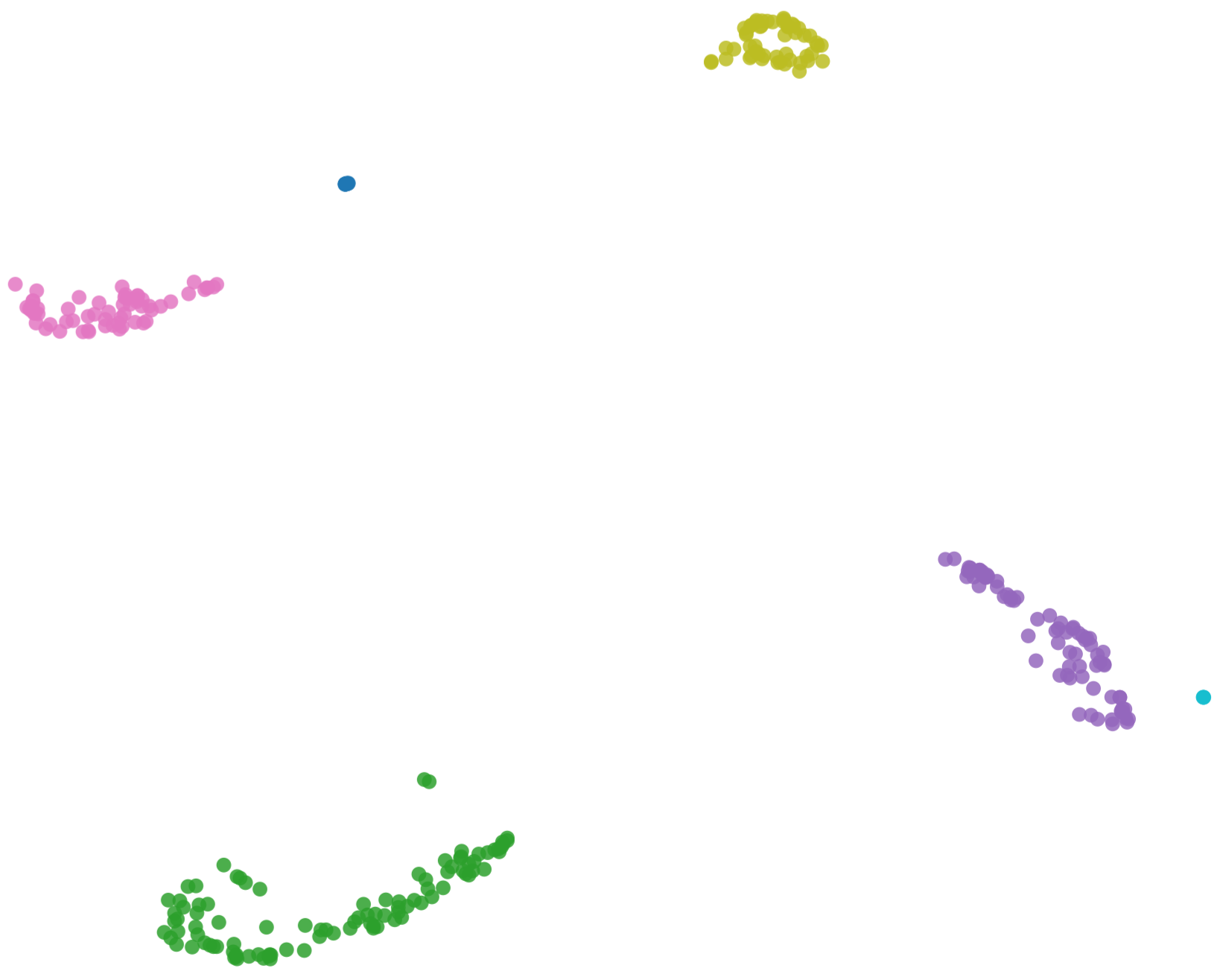}
    \caption{\scriptsize{FEC using {\rec}}}
    \label{f3d}
\end{subfigure}
\begin{subfigure}[t]{0.22\textwidth}
    \centering
    \includegraphics[width=3.5cm,height=1.1cm]{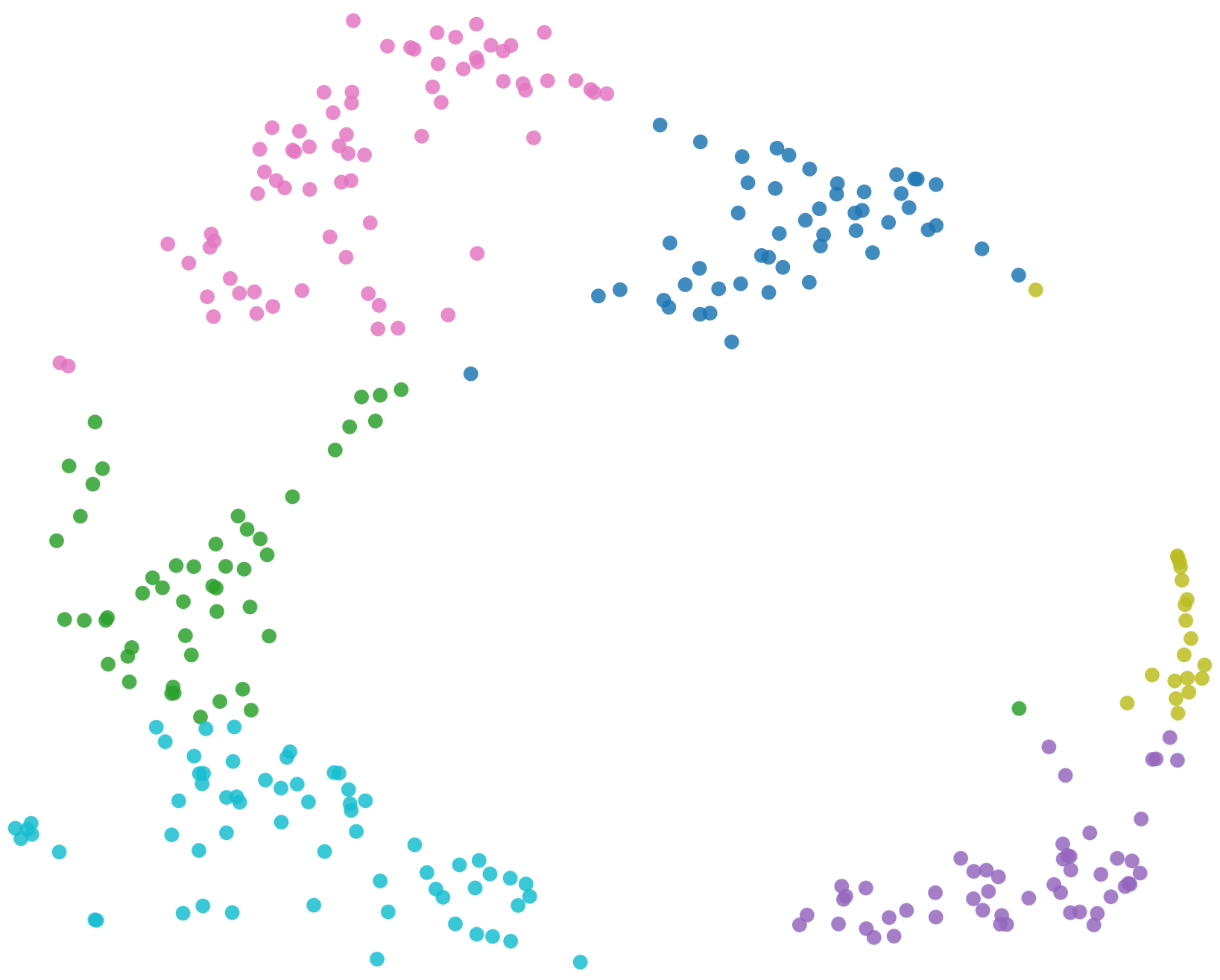}
    \caption{\scriptsize{SPONGE}}
    \label{f3c}
\end{subfigure}
\hfill
\begin{subfigure}[t]{0.22\textwidth}
    \centering
    \includegraphics[width=3.5cm,height=1.1cm]{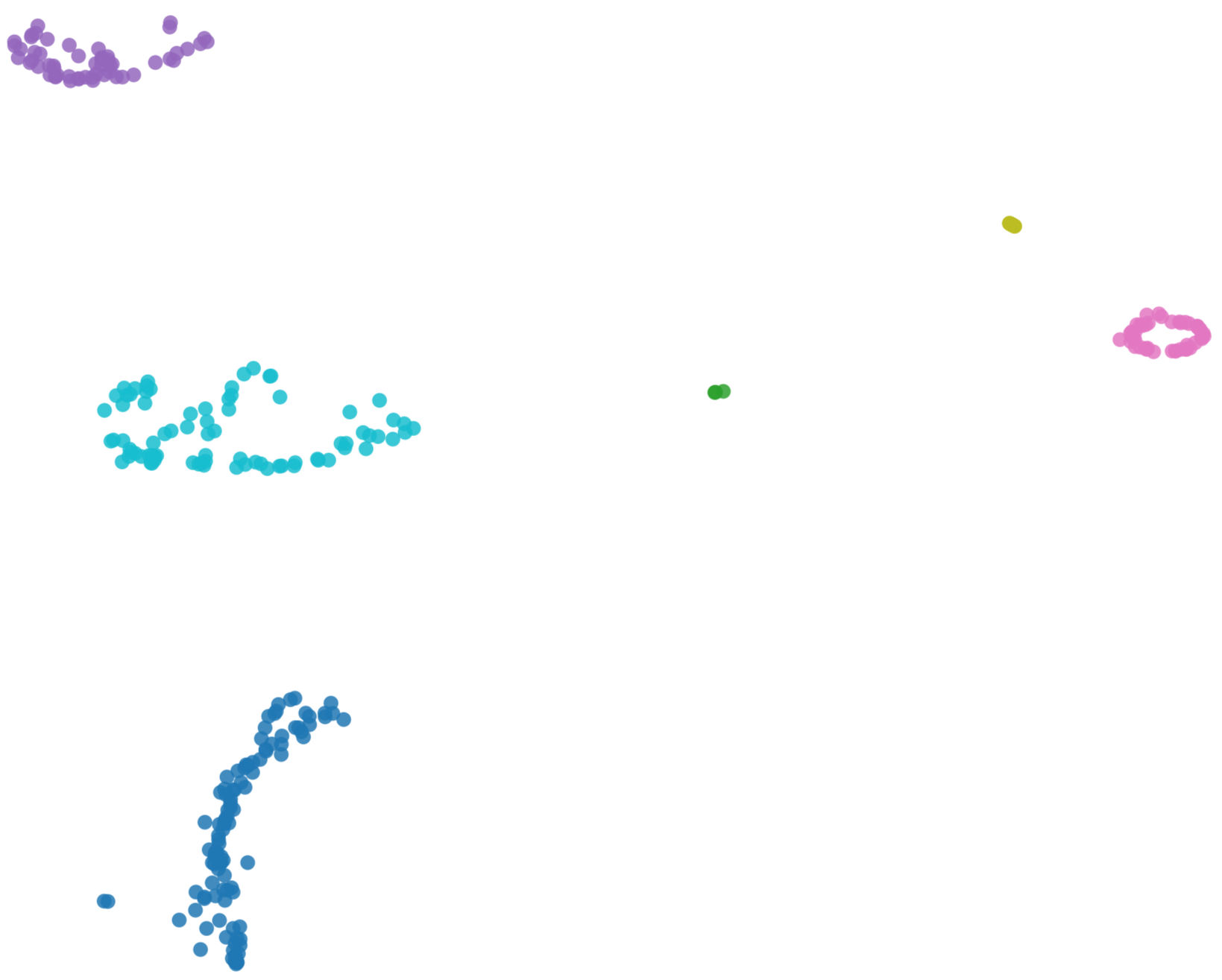}
    \caption{\scriptsize{SPONGE using {\rec}}}
    \label{f3d}
\end{subfigure}
\vspace{-4mm}
\caption{Visualization of FEC and SPONGE on Rainfall.}
\label{fig_3}
\vspace{-5mm}
\end{figure}